\documentclass[aip,rsi,reprint,graphicx,nofootinbib]{revtex4-2} 

\draft 

\usepackage{amsmath}    
\usepackage{verbatim}   
\usepackage{color}      
\usepackage{subfigure}  
\usepackage{placeins}	
\usepackage[pdftex]{graphicx} 
\usepackage{epstopdf}
\usepackage[colorlinks=true,allcolors=black]{hyperref}
\usepackage{textcomp}
\usepackage[english]{babel}
\usepackage{gensymb}
\usepackage{ulem}
\usepackage{wasysym}
\usepackage{xcolor}
\usepackage{units}
\usepackage{url}

\usepackage{breakurl}
\usepackage{etoolbox}
\appto\UrlBreaks{\do\-}
\hypersetup{
	colorlinks=true,
	linkcolor=blue,
	citecolor=blue,
	urlcolor=blue,
	breaklinks=true}

\begin{document}
	\title{Bichromatic UV detection system for atomically-resolved imaging of ions} 
	
	\author{T. Nordmann} 
	\affiliation{Physikalisch-Technische Bundesanstalt (PTB), Bundesallee 100, 38116 Braunschweig, Germany}
	\author{S. Wickenhagen} 
	\affiliation{Asphericon GmbH, Stockholmer Str. 9, 07747 Jena, Germany}
	\author{M. Dole{\v z}al} 
	\affiliation{Czech Metrology Institute (CMI), Okru{\v z}n{\' i} 31, 638 00 Brno, Czech Republic}	
	\author{T. E. Mehlst{\"a}ubler} 
	\email[]{Tanja.Mehlstaeubler@ptb.de}
	\affiliation{Physikalisch-Technische Bundesanstalt (PTB), Bundesallee 100, 38116 Braunschweig, Germany}		
	\affiliation{Institut f{\"u}r Quantenoptik, Leibniz Universit{\"a}t Hannover, Welfengarten 1, 30167 Hannover, Germany}
	\affiliation{Laboratorium f{\"u}r Nano- und Quantenengineering, Leibniz Universit{\"a}t Hannover, Schneiderberg 39, 30167 Hannover, Germany}				
	
	\date{\today}
	
		\begin{abstract}
		We present a compact and bichromatic imaging system, located outside of the vacuum chamber of a trapped ion apparatus, that collects the fluorescence of 230.6\,nm and 369.5\,nm photons simultaneously on a shared electron multiplying charge-coupled device (EMCCD) camera. The system contains two lens doublets, consisting of a sphere and an asphere. It provides a numerical aperture of 0.45 and 0.40 at 230.6\,nm and 369.5\,nm, respectively, and enables spatially resolved state detection with a large field of view of 300\,$\mu$m for long $^{115}$In$^+$/$^{172}$Yb$^+$ Coulomb crystals. Instead of diffraction limited imaging for one wavelength, the focus in this system is on simultaneous single-ion resolved imaging of both species over a large field with special attention to the deep UV wavelength (230.6\,nm) and the low scattering rate of In$^+$ ions. The introduced concept is applicable to other dual-species applications.
	\end{abstract}
	
	\pacs{}
	
	\maketitle 

	\section{Introduction}
	\label{introduction} 
	Trapped atomic ions are highly controllable quantum systems, isolated from the environment. Consequently, they are well suited for quantum information processing and for precision spectroscopy. Information about both the internal and external degrees of freedom of these quantum systems is gained from the fluorescence light emitted by the trapped ions. For this reason, the detection system is a central part for each application and a high collection efficiency and thereby fast detection directly influences the fidelity of quantum operations or the stability of frequency standards.\\
	
	In the case of a single species the most common and flexible choice is a multi-lens system outside of the vacuum chamber \cite{Pyka_2014, Alt_2002, Noek_2013, Wong_2016}. To obtain large collection efficiencies cavity assisted approaches \cite{Sterk_2012, Casabone_2015, Steiner_2013, Ballance_2017} are followed or stationary collection lenses \cite{Koo_2004}, large parabolic \cite{Chou_2017, Maiwald_2012, Alber_2017, Wang_2020} or spherical mirrors \cite{Shu_2009, Hetet_2010, Shu_2011} inside the vacuum chamber are used to collect the emitted photons. For the detection of long ion crystals a parabolic or spherical mirror is difficult because of a limited field of view of the optical system. \\
	
	Applications involving two different atomic species require achromatic imaging at often far separated wavelengths. In the deep UV spectral region the common lens materials exhibit substantially stronger chromatic dispersion than in the visible or infrared region which prevents the design of achromatic lens systems involving a deep UV wavelength in combination with a larger wavelength. Instead, fully reflective Schwarzschild objectives are used \cite{Dubielzig_2021} or modified types. One example for a modified type, which images at 280\,nm and 313\,nm, includes an achromatic lens in addition to the reflective elements and is a relatively large-sized, long working distance objective outside of the chamber \cite{Huang_2004}. Another possibility is the use of a monolithic piece of fused silica, inside the vacuum, reflectively coated with aluminum, to maintain a rigid alignment of the two Schwarzschild mirrors. Here, achromaticity is preserved by shaping the refracting surfaces of the disperse medium such that rays cross them perpendicularly \cite{Fujiyoshi_2007, Leupold_2015}. For the parallel detection at 313\,nm and 397\,nm a purely refractive system with an $NA$\,$\approx$\,0.45 consisting of five lenses inside the vacuum chamber and a dichroic beamsplitter and two compensation lenses for each species outside the chamber was developed \cite{Lo_2015}. Here the two wavelength are detected with individual electron multiplying charge-coupled device (EMCCD) cameras and PMTs.\\
	
	In this work we introduce an achromatic refractive deep UV detection setup with single-ion resolution across a field of view of 300\,$\mu$m, featuring an $NA$\,$\geq$\,0.4 for both species. The two species are imaged simultaneously onto the same EMCCD camera and the number of refractive elements is significantly lower. Automatized translation stages enable the detection of ions in all trap segments of the scalable, linear ion trap \cite{Keller_2019_PRA}. Our approach is applicable to all directly detectable dual-species quantum systems. In the field of optical clocks this applies for instance to systems such as $^{115}$In$^+$/$^{40}$Ca$^+$ \cite{Ohtsubo_2017}, $^{115}$In$^+$/$^{172}$Yb$^+$ \cite{Herschbach_2012} and $^{171}$Yb$^+$/$^{88}$Sr$^+$. In quantum information processing applications using e.g. $^{40}$Ca$^+$/$^{9}$Be$^+$ \cite{Negnevitsky_2018}, $^{138}$Ba$^+$/$^{171}$Yb$^+$ \cite{Sakrejda_2021, Brown_2016, Inlek_2017, Pino_2021}, $^{25}$Mg$^+$/$^{9}$Be$^+$ \cite{Wan_2019, Erickson_2022}, $^{40}$Ca$^+$/$^{88}$Sr$^+$ \cite{Bruzewicz_2019} and $^{43}$Ca$^+$/$^{88}$Sr$^+$ \cite{Hughes_2020} may benefit from this approach.\\
	
	This paper is organized as follows: in Sec~\ref{experimental_setup} we present the experimental apparatus and the ion species. The specifications for the systems, its design, tolerances and mounting are presented in Sec~\ref{optical_system}. In Sec.~\ref{Characterization} the detection system is characterized and Sec~\ref{conclusion} closes with a conclusion.\\
	
	\section{Experimental setup}
	\label{experimental_setup}
\begin{figure}[h!]
		\includegraphics[width=6.0cm]{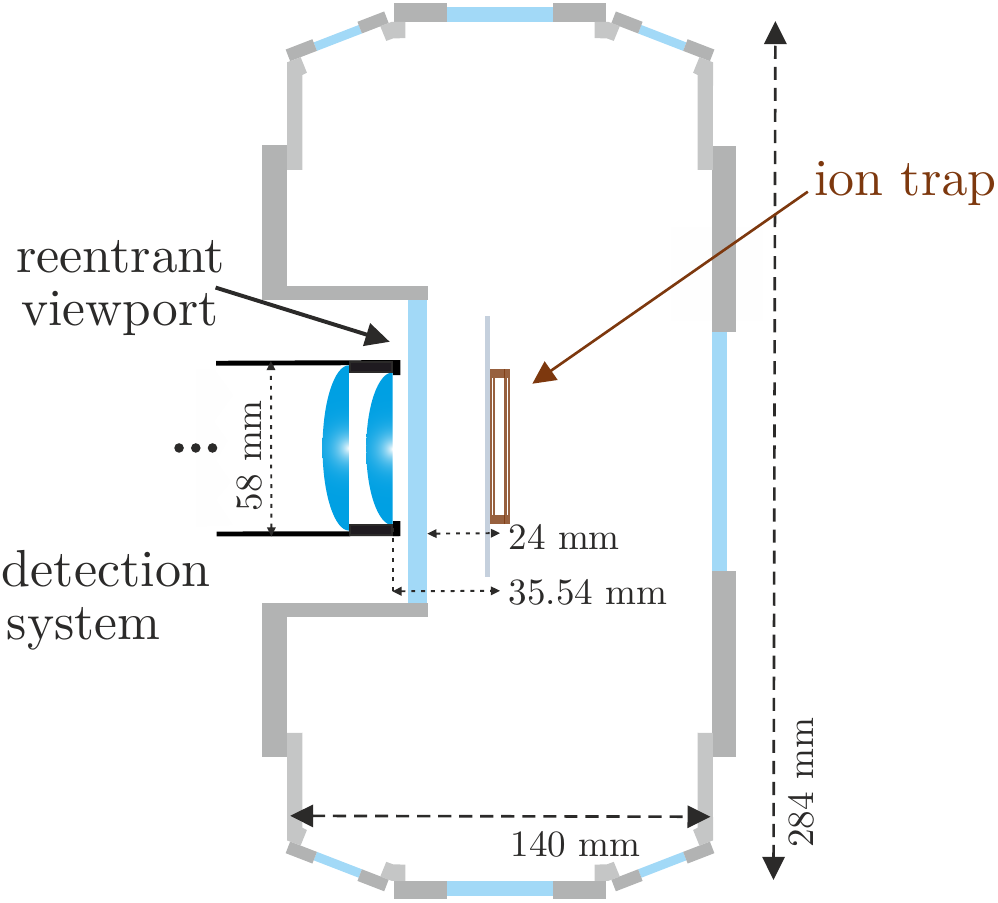}
		\caption{Drawing of the horizontal plane of the vacuum chamber with the ion trap at its center. The detection system is placed behind the 6.35\,mm thick reentrant viewport, the distance between the ions and the viewport is 24\,mm and the working distance of the detection system is 35.54\,mm.}
		\label{fig:vacuum_chamber_reentrant}
	\end{figure}

	Coulomb crystals consisting of $^{115}$In$^+$ and $^{172}$Yb$^+$ ions are trapped in a linear segmented Paul trap which is placed in an octahedral vacuum chamber. To enable a large $NA$ the vacuum system features a 6.35\,mm thick reentrant viewport with a distance of 24\,mm between the ions and the viewport. The working distance of the imaging system is 35.54\,mm. A drawing of the horizontal plane of the chamber with the reentrant viewport and the dimensions is shown in Fig.\ref{fig:vacuum_chamber_reentrant}.
	
	In$^+$ ions feature a narrow $^1S_0$~$\leftrightarrow$~$^3P_1$ intercombination line at a wavelength of 230.6\,nm. The low scattering rate $\Gamma$~=~2$\pi$~$\times$~360\,kHz and the deep UV wavelength make fluorescence detection challenging and require a high collection efficiency. The demands for detecting Yb$^+$ ions on the $^2S_{1/2}$~$\leftrightarrow$~$^2P_{1/2}$ transition at 369.5\,nm with a scattering rate $\Gamma$~=~2$\pi$~$\times$~19.6\,MHz are somewhat more relaxed.\\
	
	\section{Optical system}
	\label{optical_system}
	\subsection{System specifications}
	\label{system_specifications}
	The target for the detection system is the simultaneous imaging of the fluorescence of $^{115}$In$^+$ and $^{172}$Yb$^+$ ions on a shared EMCCD camera, thereby providing spatially resolved state detection of both species for spectroscopy. Another requirement is the collection of the fluorescence of both species on individual photomultiplier tubes (PMT).\\
	
	The design goal is a wide field of view of 300\,$\mu$m. The smallest possible working distance of the system is 31\,mm. The spatial limitations of the reentrant viewport allow a lens diameter of 2\," enabling an $NA$ of 0.4-0.5. The targeted magnification is between 12 and 16. The lower bound is given by the requirement to resolve the individual ions on the 16\,$\mu$m sized pixels of the EMCCD camera\footnote{Andor, iXon Ultra 897}. In a 20-ion chain, which is trapped at a radial secular frequency of 2$\pi$\,$\cdot$\,1\,MHz (for Yb$^+$) the maximum axial secular frequency at which the ions still form a linear chain is about 2$\pi$\,$\cdot$\,100\,kHz (for Yb$^+$). The corresponding distance of the inner ions is 4.6\,$\mu$m and results in a distance of 3.5\,pixels for the lower bound of the magnification. The upper bound ensures that the full field of view is imaged on the 8\,mm sized camera chip.	Ideally, each ion is imaged to a single pixel to obtain the maximum signal-to-noise ratio. The targets of simultaneous imaging of both UV wavelengths and single-ion resolution are significantly more important than the noise properties. Given the minimal distance of 3.5\,pixels between two ions that needs to be resolved, imaging each ion to 2x2 or 3x3 pixels is also sufficient.\\
	
	In the following, we describe our approach for the design optimization. The evaluation criterion is the fraction of enclosed energy $f_\mathrm{ee}$ in a given radius $r_\mathrm{c}$ from centroid.\\
	
	A diffraction limited system can be characterized by its Airy disk with a full width at half maximum (FWHM) diameter of \cite{Rayleigh_1879, Leica_2014}:
	\begin{align}
		\label{eq:FWHM_Airy}
		\delta_{\mathrm{A}}=\frac{0.51 \cdot \lambda }{NA}, 
	\end{align}
	where $\lambda$ is the wavelength of the imaging light. The fraction of the energy enclosed in the radius $r_{\mathrm{A}}$\,=\,$\delta_{\mathrm{A}}$/2 is $f_\mathrm{ee}$($r_{\mathrm{A}}$)\,=\,0.76.\\
	
	The radius $r_{\mathrm{A}}$  is used to classify the system via simulations of the fraction of enclosed energy $f_\mathrm{ee}$ in a given radius $r_\mathrm{c}$. Since the simulations contain the magnification $M$, Eq.~\ref{eq:FWHM_Airy} needs to be corrected for $M$ in order to compare the values, which results in:
	\begin{align}
		\label{eq:r_Airy_M}
		r_{\mathrm{A}}=\frac{0.51 \cdot \lambda \cdot M }{2 NA}.
	\end{align}
	\\
	
	\subsection{Design}
	\label{design}
\begin{center}
		\begin{figure}[h!]
			\includegraphics[width=14cm]{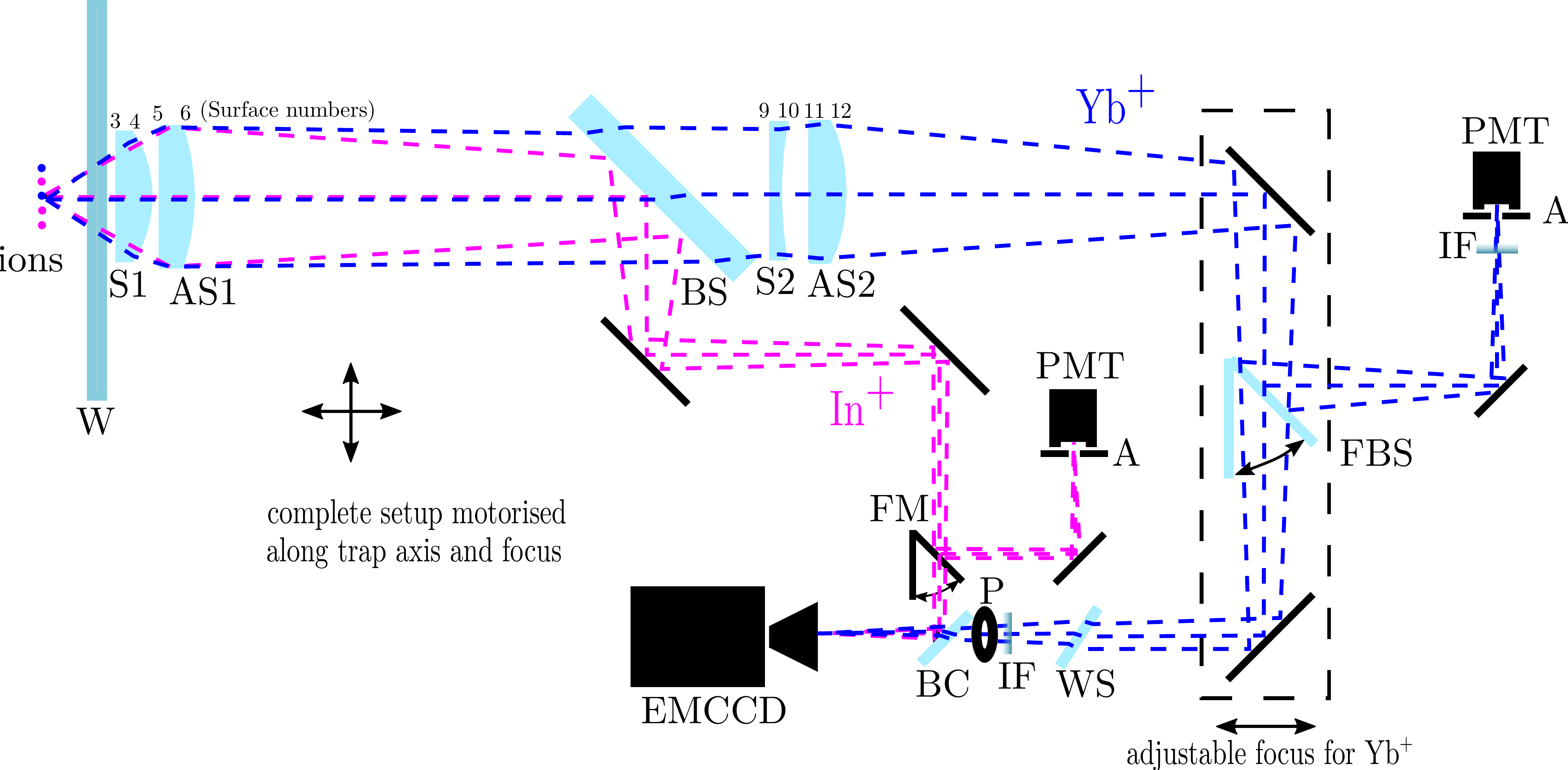}
			\caption{Setup of the bichromatic detection system. The In$^+$ and Yb$^+$ fluorescence at 230.6\,nm and 369.5\,nm are collected by a lens doublet (S1 and AS1) and are then separated by a dichroic beamsplitter (BS). The In$^+$ fluorescence is reflected, while the Yb$^+$ fluorescence is transmitted and passes through a second lens doublet (S2 and AS2) and is guided over two mirrors to be overlapped with the In$^+$ fluorescence again and both wavelengths are simultaneously imaged onto the EMCCD camera. Abbreviations: vacuum window (W), spherical lens (S1/S2), aspherical lens (AS1/AS2), beam combiner (BC), flippable mirror/beamsplitter (FM/FBS), wedged substrate (WS), pinhole (P), slit aperture (A), interference filter (IF). The small numbers are used to label the surfaces.}
			\label{fig:detection_setup}	
		\end{figure}
	\end{center}

	Figure~\ref{fig:detection_setup} shows a schematic drawing of the imaging system. Fluorescence light of both ion species passes through the 6.35\,mm thick vacuum window (W) and is collected by the first lens doublet consisting of a sphere (S1) and an asphere (AS1). The next optical element is a beam splitter (BS), which reflects the In$^+$ wavelength but transmits the Yb$^+$ wavelength (cutoff wavelength for reflectance: 245\,nm). The Yb$^+$ image is then corrected by a second lens doublet consisting of a sphere (S2) and an asphere (AS2). Afterwards it is reflected by two mirrors and transmitted through a 1\,$^{\circ}$ wedged substrate (WS). Finally, a beam combiner (BC) combines the In$^+$ and Yb$^+$ fluorescence again on the EMCCD camera. The wedged substrate compensates aberrations induced by the beamsplitter and the beamcombiner. Omitting the WS would lead to a $\approx$\,6 times broader image and distortion. All lenses have a anti-reflection coating optimized for both wavelengths (curved surfaces: R$<$2\% at 230nm and R$<$2.5\% at 370nm, plane surfaces of S1 and AS1: R$<$1\% at 230nm and R$<$1\% at 370nm, plane surfaces of S2 and AS2: R$<$0.2\% at 370nm).\\
	\onecolumngrid
	\twocolumngrid
	The system is fully motorized: besides the motors to adjust the focus and axial position to image ions in all eight segments of the trap \cite{Keller_2019_PRA}, the system provides two automated flip mirrors to send the fluorescence of both species on photomultiplier tubes (PMTs). The Yb$^+$ fluorescence is filtered by interference filters (IF) with a transmission $T$\,=\,0.9 and a bandwidth $BW$\,=\,$\pm$\,6\,nm in front of the PMT and EMCCD each, to protect those from background stray light. For the In$^+$ beam path no filter is needed since the coating of the substrates is only reflective for wavelengths within a 40-60\,nm band around the In$^+$ intercombination line at 230.6\,nm and the PMT is solar blind\footnote{CsTe cathode with sensitivity range: 160\,nm-320\,nm, QE\,=\,30\,\% at 230nm, Hamamatsu R7154}. The two mirros behind the second lens doublet are mounted on a shared translations stage to adjust minor focus differences in the two beam paths individually. These two mirrors are also the only mirrors that can be tilted in two directions to control the position of the Yb$^+$ image on the camera independently of the In$^+$ image. In general, the number of degrees of freedom was kept as low as possible to achieve the best possible stability and simplest possible adjustment of the complex system. In front of both PMTs a slit aperture (A) is mounted to protect them from stray light e.g.  scattered light from the trap electrodes.\\
	
	With both wavelengths in the UV, the number of optical elements was kept to a minimum, therefore aspheric lenses were chosen. Because of the narrower linewidth and the lower wavelengths of the fluorescence of the In$^+$ compared to the Yb$^+$ ions, the focus for the optimization was set on lowest aberrations at 230.6\,nm. For this reason, as few as possible transmittive elements are used in the deep UV light path. Furthermore, dichroic beamsplitters are easier to realize when the longer wavelength is transmitted and the shorter reflected, respectively \cite{Laseroptik}.\\
	
	The first design idea was to use two biaspheric lenses instead of lens doublets, but the relative centering of two aspheric lens surfaces could not reach the necessary tolerances. Lenses with one planar surface can be machined with much higher precision.\\
	
	The system was optimized for a field of view of $d$\,=\,300\,$\mu$m and simulated at three different positions in the objective plane: at 0\,$\mu$m, 106\,$\mu$m and 150\,$\mu$m off the optical axis. The point spread function (PSF) was used as the optimization criterion and a fraction of enclosed energy $f_\mathrm{ee}$\,=\,90\% on a single 16\,$\mu$m\,$\times$\,16\,$\mu$m pixel ($r_\mathrm{c}$\,=\,8$\mu$m) for 230.6\,nm and on 2\,$\times$\,2 pixels ($r_\mathrm{c}$\,=\,16$\mu$m) for 369.5\,nm was targeted, respectively. Table~\ref{tab:surface_properties} lists the result of the optimization. The numbering of the surfaces is indicated in Fig.\ref{fig:detection_setup}. The front surfaces of all lenses, surfaces 3, 5, 9 and 11, are plane. Surfaces 6 and 12 are aspheric and surfaces 4 and 10 are spheric. The optimized system has an $NA$\,=\,0.45 and a magnification $M$\,=\,12 for 230.6\,nm and $NA$\,=\,0.40 and $M$\,=\,10 for 369.5\,nm. The $NA$ is limited by the diameter of the second lens for the 230.6\,nm path and by the fourth lens for the 369.5\,nm path. A wedged substrate (WS) was introduced to compensate aberrations which are introduced by the beamsplitter (BS) in the 369.5\,nm path. These two components and the vacuum window (W) have not been considered in the tolerancing, described in Sec.~\ref{tolerances}, as this would have exceeded
	the complexity of the tolerancing. In order to achieve a high collection efficiency on the PMT and a possibly aberration free image on the camera, an aperture was included on the way to the camera to reduce the $NA$ of the Yb$^+$ camera path from 0.4 to $\approx$\,0.25.
	\onecolumngrid
	\begin{center}
		\begin{table}[htbp!]
			\caption{Surfaces as given in the Zemax lens data editor. The numbering of the surfaces is consistent with Fig.~\ref{fig:detection_setup}. The coefficients of the aspheric surfaces 6 are k=-3.7, A$_4$=1.065~$\times$~10$^{-6}$ and A$_6$=-8.995~$\times$~10$^{-10}$ and for surface 12 k=0.83, A$_4$=-3.956~$\times$~10$^{-7}$ and A$_6$=-1.714~$\times$~10$^{-10}$}
			\label{tab:surface_properties}
			\begin{center}
				\begin{ruledtabular}
					\begin{tabular}{lllllll} 
						Lens &Surface& Type &Radius (mm) & Thickness (mm) & Diameter (mm) &Glass\\
						\toprule
						S1 & 3 & Sphere &Infinity &15.00 &44.00 &Corning 7980\\
						S1 & 4 &Sphere &-34.59 &5.00 &44.00 &\\
						AS1 & 5 & Sphere &Infinity &15.00 &50.80 &Corning 7980\\
						AS1 & 6 & Even asphere &-57.708 &95.00 &50.80 &\\
						S2 & 9 & Sphere &Infinity &5.00 &48.80 &Corning 7980\\
						S2 & 10 & Sphere &53.68 &10.00 &48.80 &\\
						AS2 & 11 & Sphere &Infinity &15.00 &50.80 &Corning 7980\\
						AS2 & 12 & Even asphere &-51.60 &197.825 &50.80 &\\
					\end{tabular}
				\end{ruledtabular}
			\end{center}
		\end{table}
	\end{center}
	\twocolumngrid

\begin{figure}[hb!]
		\includegraphics[width=9.0cm]{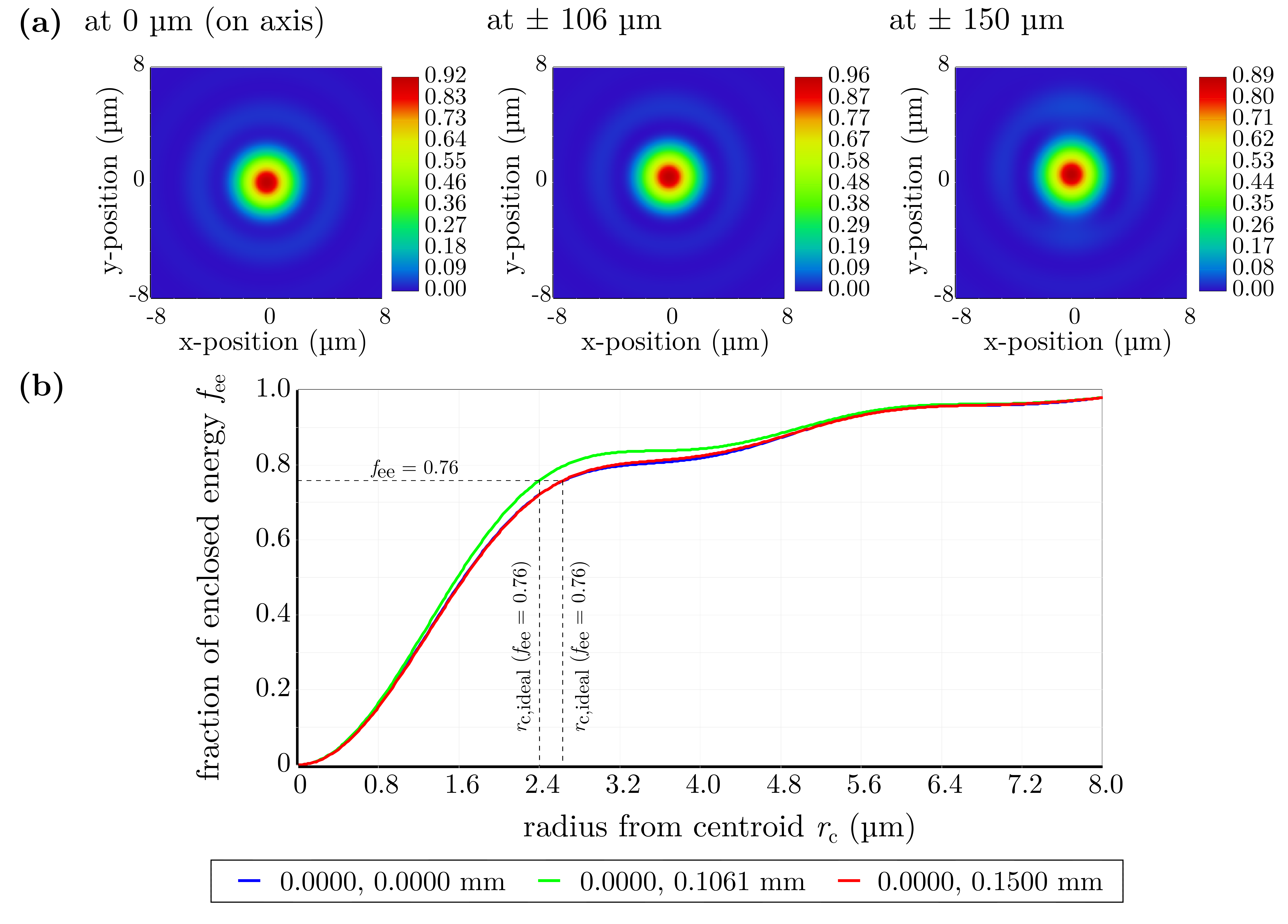}
		\caption{2D point spread function (PSF) (a) and fraction of enclosed energy $f_\mathrm{ee}$ versus the radius from centroid $r_\mathrm{c}$ (b) of the optimized 230.6\,nm path at three positions in the objective plane: 0\,$\mu$m, $\pm$106\,$\mu$m and $\pm$150\,$\mu$m (off the optical axis).}
		\label{fig:spot_diagram_PSF_230nm}
	\end{figure}

\begin{figure}[h!]
		\includegraphics[width=9.0cm]{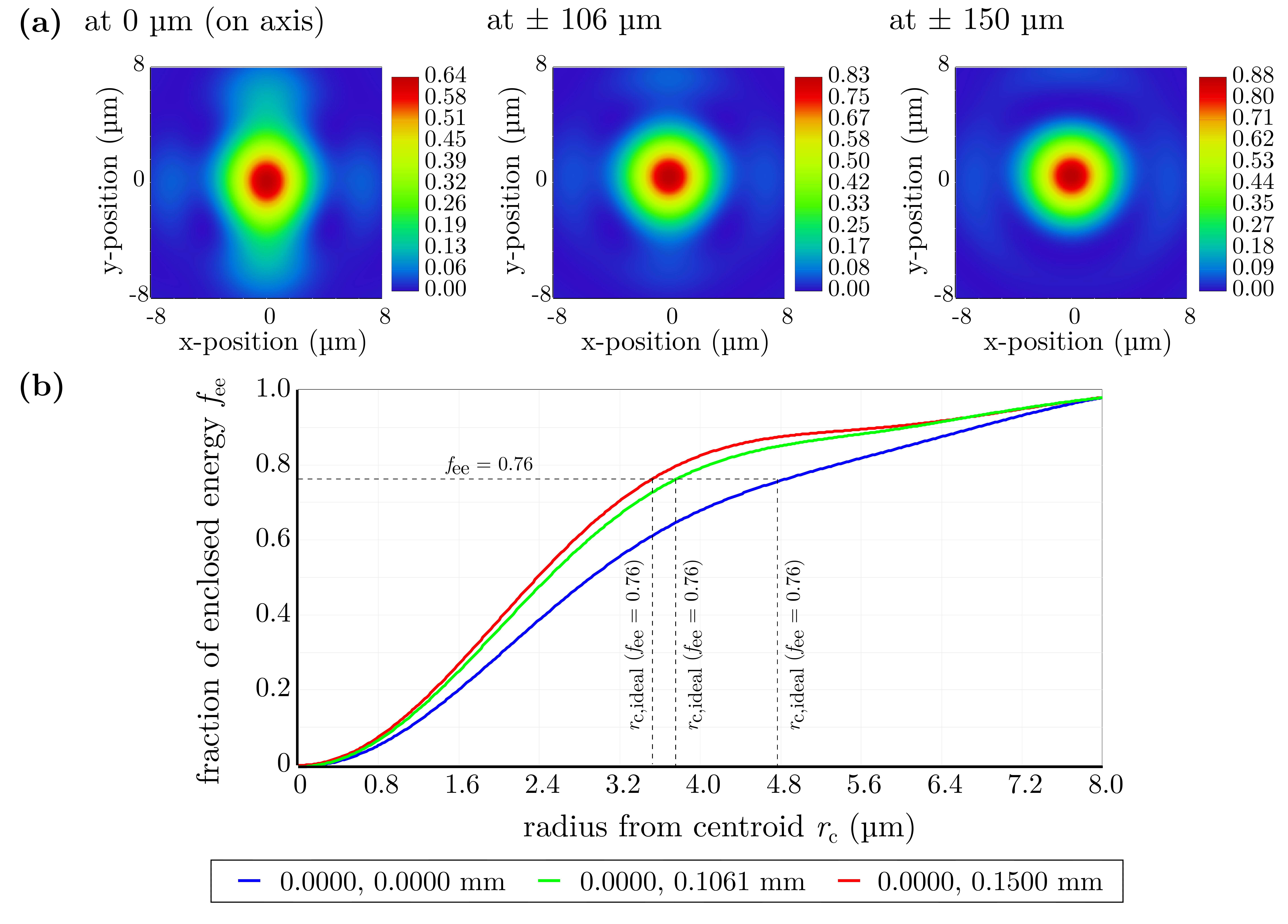}
		\caption{2D point spread function (PSF) (a) and fraction of enclosed energy $f_\mathrm{ee}$ versus the radius from centroid $r_\mathrm{c}$ (b) of the optimized 369.5\,nm path at three positions in the objective plane: 0\,$\mu$m, $\pm$106\,$\mu$m and $\pm$150\,$\mu$m (off the optical axis).}
		\label{fig:spot_diagram_PSF_369nm}
	\end{figure}

	Figure~\ref{fig:spot_diagram_PSF_230nm} shows the 2D point spread function (PSF) and the fraction of enclosed energy $f_\mathrm{ee}$ of the 230.6\,nm path for the three optimization points in the objective plane. Fig.~\ref{fig:spot_diagram_PSF_369nm} illustrated the same quantities for the 369.5\,nm path. It is noticeable that the imaging quality of the 369.5\,nm beam path is significantly better for the off-axis positions. The on-axis PSF shows a pale vertical and horizontal stripe.\\
	
	\begin{table}[hbp!]
		\caption{Comparison of the theoretical diffraction-limited radius $r_{\mathrm{A}}$\,=\,$\delta_{\mathrm{A}}$/2 at which the fraction of enclosed energy $f_{\mathrm{ee}}$($r_{\mathrm{A}}$)\,=\,0.76 with the radius from centroid  $r_{\mathrm{c,ideal}}$ at which $f_{\mathrm{ee}}$\,=\,0.76 for the ideal system according to the simulations in Fig.~\ref{fig:spot_diagram_PSF_230nm} and Fig.~\ref{fig:spot_diagram_PSF_369nm}.}
		\label{tab:comparison_rA_rc}
		\begin{center}
			\begin{ruledtabular}
				\begin{tabular}{lll} 
					& $r_{\mathrm{A}}$ &  $r_\mathrm{c,ideal}$($f_\mathrm{ee}$\,=\,0.76) \\
					& ($\mu$m) &  ($\mu$m) \\
					\toprule
					230.6\,nm &  1.57  &2.4-2.6\\
					369.5\,nm &  2.36  &3.5-4.8\\
				\end{tabular}
			\end{ruledtabular}
		\end{center}
	\end{table}
	
	Table~\ref{tab:comparison_rA_rc} compares the radius $r_{\mathrm{A}}$\,=\,$\delta_{\mathrm{A}}$/2, at which a diffraction limited system reaches a fraction of enclosed energy $f_{\mathrm{ee}}$($r_{\mathrm{A}}$)\,=\,0.76, with the radius $r_{\mathrm{c,ideal}}$($f_{\mathrm{ee}}$\,=\,0.76) corresponding to the simulations of the optimized system. The simulations of the optimized system are shown in Fig.~\ref{fig:spot_diagram_PSF_230nm} and Fig.~\ref{fig:spot_diagram_PSF_369nm} where $f_{\mathrm{ee}}$\,=\,0.76 and the respective values for $r_{\mathrm{c,ideal}}$ are indicated. For both wavelengths the design radius $r_{\mathrm{c,ideal}}$($f_{\mathrm{ee}}$\,=\,0.76) is a factor 1.5-2 larger than the diffraction limited $r_{\mathrm{A}}$. This is due to the cost of keeping a large field of view of $d$\,=\,300\,$\mu$m. Still imaging both species to a single pixel of a size 16\,$\mu$m\,$\times$\,16\,$\mu$m is possible, as $f_\mathrm{ee,1p}$($r_\mathrm{c,1p}$\,=\,8\,$\mu$m)\,=\,0.98 for both wavelengths. However, all given values discussed so far do not include any manufacturing tolerances.\\
	
	\subsection{Tolerancing}
	\label{tolerances}
	The performance of the system discussed in section~\ref{design} is idealistic. The realistic performance will be lower due to manufacturing tolerances of the lens surfaces and the mounting.\\
	
	In the following, the tolerancing is performed for the two doublet lens systems. All other components are assumed to be ideal. Table~\ref{tab:tolerances_center_thickness_and_radii} summarizes realistic tolerances for the center thickness and the radius of the front and back side of the four lenses, given by the manufacturer. With these tolerances the encircled energy for a radius corresponding to one pixel reduces slightly to $f_\mathrm{ee,1p}$($r_\mathrm{c}$\,=\,8\,$\mu$m)\,=\,0.95 for both wavelengths. \\
	
	\begin{table}[htbp!]
		\caption{Tolerances for center thickness and radii of the lenses. The sagitta deviation is the peak-to-valley deviation between the 'best-fit sphere/plane' and the target surface within the free aperture. Specifications of the sagitta deviation originate from interferometric tests with probe glasses and are given in values of interference fringes (fr), here at a reference wavelength of $\lambda_\mathrm{ref}$\,=\,546\,nm, so 1\,fringe (fr)\,=\,$\lambda_\mathrm{ref}$/2\,=\,273\,nm. For the curved back side of the lenses an absolute radius tolerance is given in addition to the sagitta deviation. The sagitta deviation and absolute radius tolerance can be converted to each other.}
		\label{tab:tolerances_center_thickness_and_radii}
		\begin{center}
			\begin{ruledtabular}
				\begin{tabular}{llll} 
					Lens & Sagitta deviation &  Radius tolerance & Center thickness \\
					& front/back side (fr) &  back side ($\mu$m) & tolerance ($\mu$m)\\
					\toprule
					S1 & 1/1  &$\pm$1 &$\pm$50\\
					AS1 & 1/1  & $\pm$3 &$\pm$50\\
					S2 & 1/1  &$\pm$3 &$\pm$50\\
					AS2 & 1/1  &$\pm$3 &$\pm$20\\
				\end{tabular}
			\end{ruledtabular}
		\end{center}
	\end{table}
	The four lenses form two doublets (S1+AS1 and S2 and AS2), each set in its own mount. More details about the mounting of the lenses are given in section~\ref{mounting}. In Tab.~\ref{tab:tolerances_tilts_and_decentering} tolerances for the mounting of the lenses are given, which already represent the limits of manufacturing feasibility. \\
	\begin{table}[htbp!]
		\caption{Tolerances for tilts and decentering of the lenses.}
		\label{tab:tolerances_tilts_and_decentering}
		\begin{center}
			\begin{ruledtabular}
				\begin{tabular}{ll} 
					Tolerance & Value \\
					\toprule
					Distance of the plane surfaces within one mount &$\pm$ 30\,$\mu$m\\
					Decentering of mount S1+AS1 &$\pm$50\,$\mu$m\\
					Decentering of mount S2+AS2 &$\pm$50\,$\mu$m\\
					Tilt of mount S1+AS1 &$\pm$2\,arcmin\\
					Tilt of mount S2+AS2 &$\pm$1\,arcmin\\
					Tilting of the plan support for S1 &$\pm$2\,arcmin\\
					Tilting of the plan support for AS1 &$\pm$2\,arcmin\\
					Tilting of the plan support for S2 &$\pm$2\,arcmin\\
					Tilting of the plan support for AS2 &$\pm$2\,arcmin\\
					Tilting front to back S1 &$\pm$2\,arcmin\\
					Tilting front to back AS1 &$\pm$2\,arcmin\\
					Tilting front to back S2 &$\pm$2\,arcmin\\
					Tilting front to back AS2 &$\pm$2\,arcmin\\
				\end{tabular}
			\end{ruledtabular}
		\end{center}
	\end{table}
	
\begin{figure}[h!]
		\includegraphics[width=8.5cm]{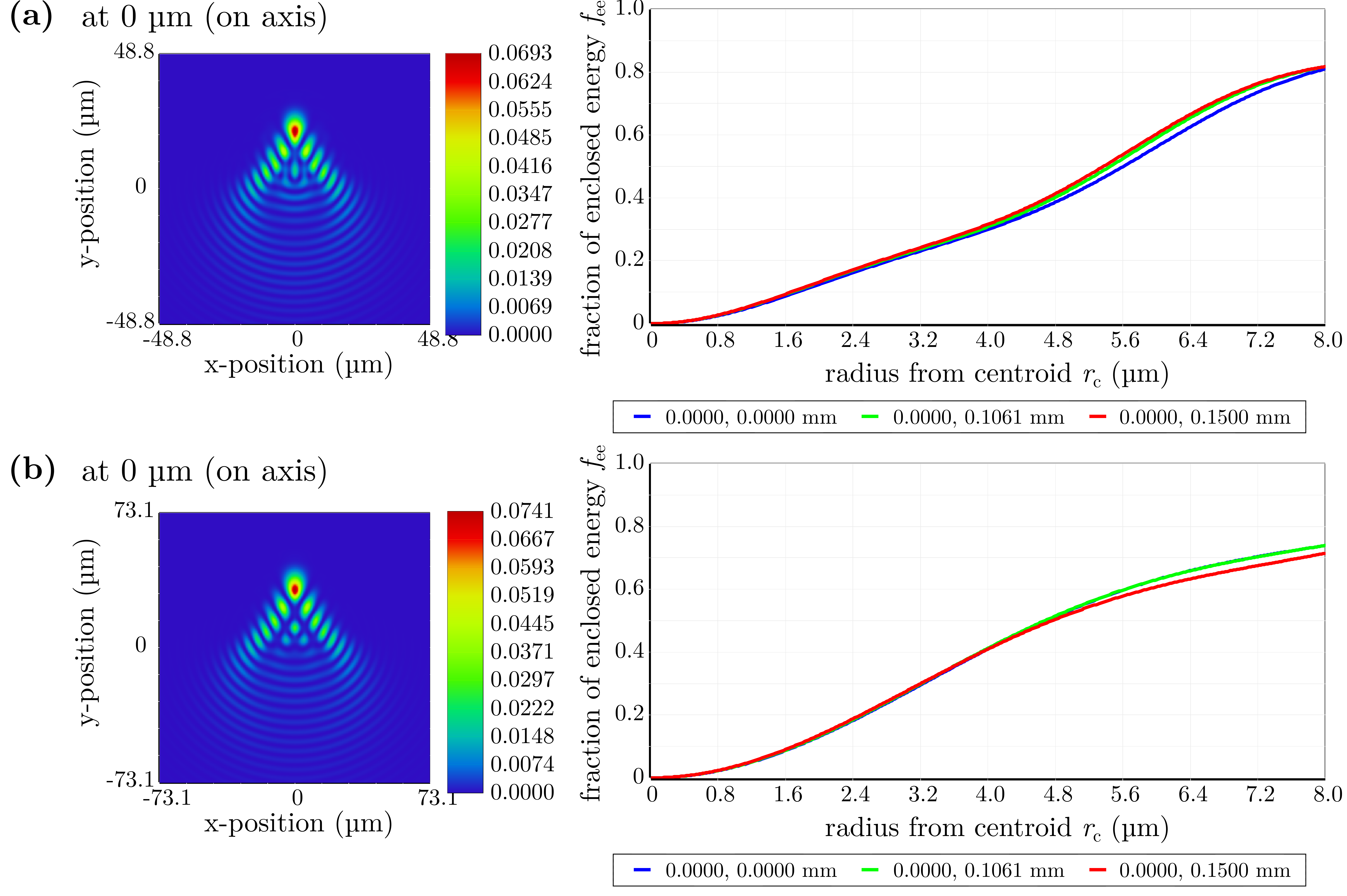}
		\caption{Worst-case on-axis 2D point spread function (PSF) and fraction of enclosed energy of the 230.6\,nm (a) and 369.5\,nm (b) path considering the tolerances for the center thicknesses, radii, tilts and decentering of the lenses and assuming all other parts to be perfect. The 2D PSFs for the two other positions are similar to the on-axis position.}
		\label{fig:PSF_with_tolerances}
	\end{figure}

	With these tolerances $f_\mathrm{ee,1p}$\,=\,0.8 for the 230.6\,nm path and $f_\mathrm{ee,1p}$\,=\,0.7 for the 369.5\,nm path in worst case as shown in  Fig.~\ref{fig:PSF_with_tolerances}. Here, worst case means the worst result out of 4000 combinations of all mentioned tolerances, shown in Table~\ref{tab:tolerances_center_thickness_and_radii} and \ref{tab:tolerances_tilts_and_decentering}. This means the expected fluorescence yield on one pixel is reduced to 80\% and 70\,\% at 230.6\,nm and 369.5\,nm, respectively. \\
	
	In addition to the tolerances mentioned above, surface form tolerances must also be taken into account. These are given in Tab.\ref{tab:surface_form_tolerances} and reduce the fluorescence enclosed in a radius $r_\mathrm{c,1p}$\,=\,8\,$\mu$m by another 8\,\%, resulting in $f_\mathrm{ee,1p}$\,$\approx$\,0.7 and $f_\mathrm{ee,1p}$\,$\approx$\,0.6 for the 230.6\,nm and 369.5\,nm, respectively, which are referred to below as $f_\mathrm{ee,tol}$. As mentioned in Sec.~\ref{design}, the beamsplitter (BS) and wedged substrate (WS) in the 369.5\,nm path, as well as the vacuum window (W), are not considered in the tolerancing process.\\
	
	\begin{table}[htbp!]
		\caption{Surface form tolerances of the lenses. The specified values refer to a reference wavelength of 546\,nm, so 1\,fringe (fr)\,=\,273\,nm.}
		\label{tab:surface_form_tolerances}
		\begin{center}
			\begin{ruledtabular}
				\begin{tabular}{lllll} 
					Lens & Irregularity & RMSi front & Irregularity& RMSi back \\
					& front side (fr)& side (nm)& back side (fr)& side (nm) \\
					\toprule
					S1 &0.5 & $<$20 &0.7 &$<$30\\
					AS1 &0.5 & $<$25 &0.5 &$<$25\\
					S2 &0.7 & $<$30 &0.8 &$<$35\\
					AS1 &1.0 & $<$40 &0.8 &$<$35\\
				\end{tabular}
			\end{ruledtabular}
		\end{center}
	\end{table}

	\subsection{Mechanical mounting}
	\label{mounting}
\begin{figure}[h!]
		\includegraphics[width=6.0cm]{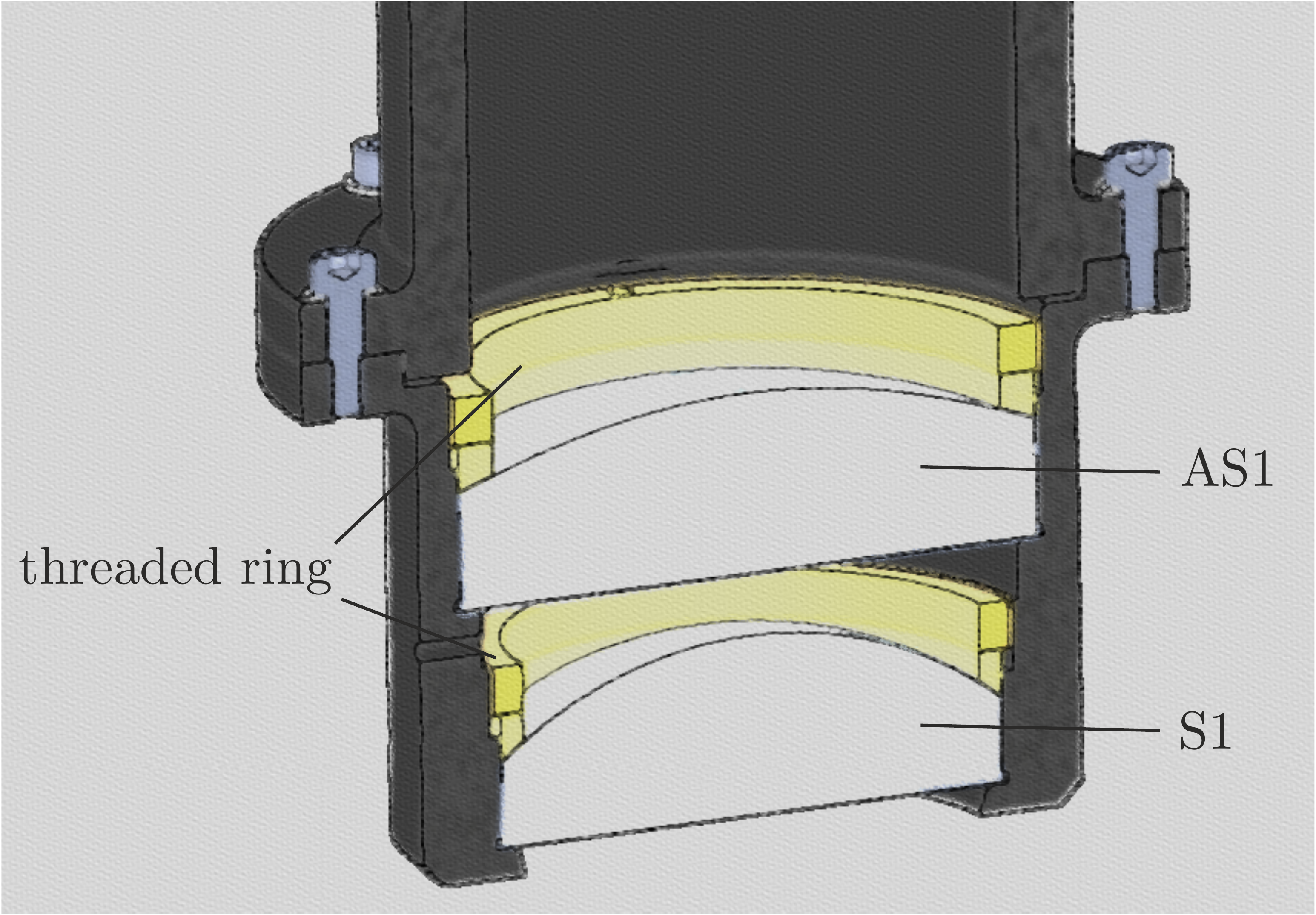}
		\caption{Mounting of the first lens doublet. The lenses are enclosed in an aluminum tube that features two two precisely machined steps that touch the plane front of the two lenses (S1 and AS1). The lenses are held in place by threaded rings (yellow).}
		\label{fig:cut_lens_mount}
	\end{figure}

	The total weight of all four lenses is $\approx$\,160\,g. For the mounting, the distance of the two lenses within one doublet and a possible tilt between the lenses are most critical. The requirement for the distance between the two doublets is lower. Both lens doublets are mounted monolithically inside a black anodized aluminum tube according to the scheme shown in Fig.~\ref{fig:cut_lens_mount}. The distance between the two steps holding the plane surfaces is machined with a tolerance of $\pm$10\,$\mu$m and a flatness of 3-4\,$\mu$m, which is better than the required tolerances stated in Table~\ref{tab:tolerances_tilts_and_decentering}.\\
	
\begin{figure}[h!]
		\includegraphics[width=9.0cm]{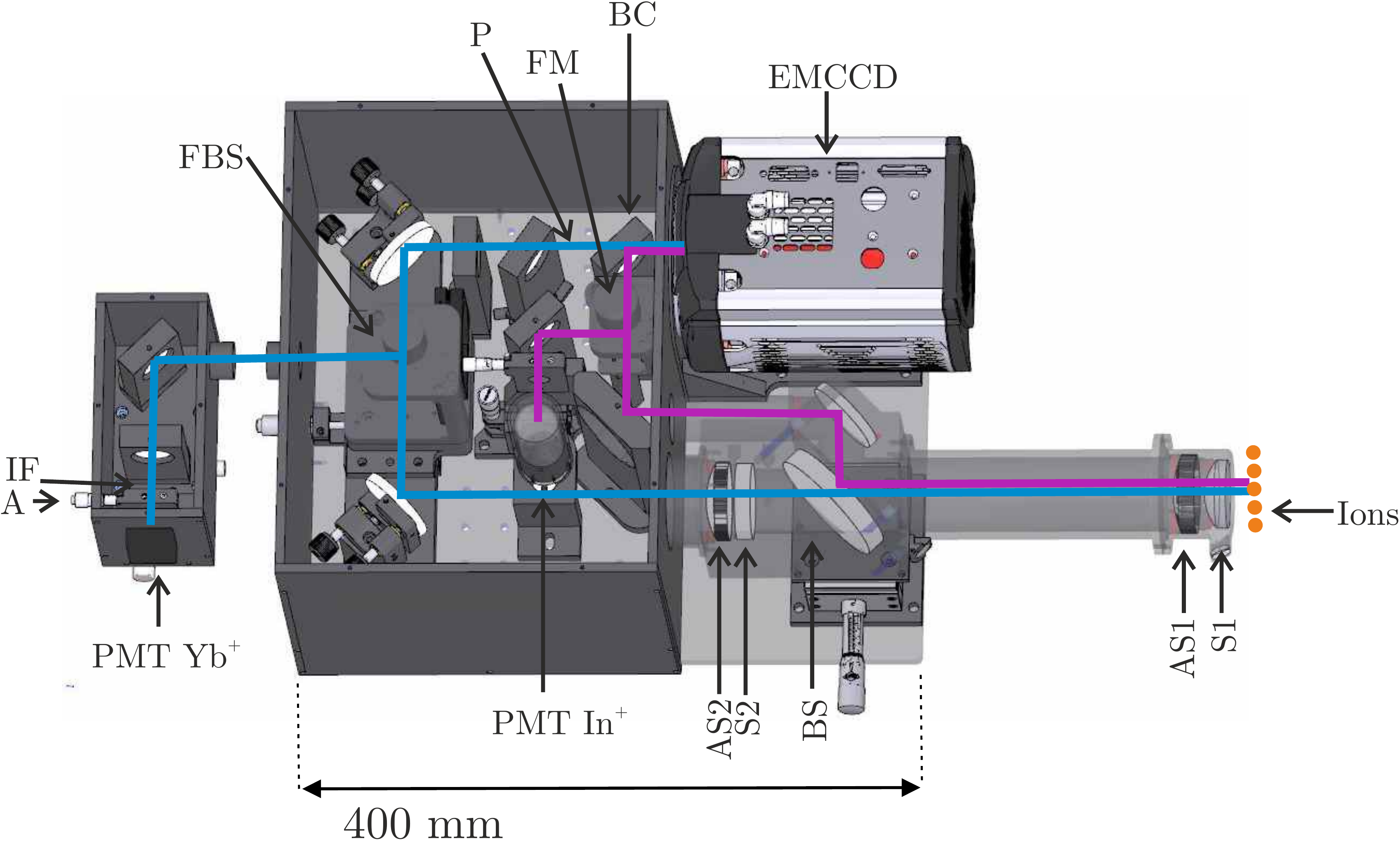}
		\caption{Mounting of all optical components. Most components are placed on an aluminum base plate inside a light-tight box. All components which require precise adjustment are equipped with fine adjustment screws. With the exception of the degrees of freedom that affect the solar-blind In$^+$ PMT, all adjustments can be made with the box closed.}
		\label{fig:detection_setup_with_mechanics}
	\end{figure}

	An overview of the mounting of the whole system is shown in Fig.~\ref{fig:detection_setup_with_mechanics}. The number of adjustable degrees of freedom is reduced to a minimum in order to keep the complexity of the setup as low as possible. Apart from the Yb$^+$ PMT, all parts are mounted on a 300\,mm$\times$400\,mm aluminum plate with a thickness of 10\,mm. The two mounted lens doublets and the beam splitter (BS) are connected via a black anodized aluminum tube and mounted on a height adjustable table. Both beampaths are guided into a light tight box where the rest of the optical elements is placed. The size and the height of the slit aperture in front of the In$^+$ PMT is adjustable with a high-precision screw. For the Yb$^+$ path, the last two mirrors are placed on a shared translation stage to adjust the focus of the two beam paths individually and the wedged substrate (WS) is fixed on a rotational stage. Both stages can be adjusted from outside the box. Motorized flip mirror holders\footnote{Owis: KSHM 40-LI-MDS and KSHM 90-LI-MDS} with a repeatability of less than 100\,$\mu$rad are used to guide the fluorescence on either the PMT or the EMCCD camera in case of the In$^+$ path (FM) and on both in equal proportions or only on the EMCCD in case of the Yb$^+$ path (FBS). The Yb$^+$ PMT is mounted on the wall of the small box and connected in a light-tight construction with the interference filter (IF) and the slit aperture (A). The small box is mounted on a height adjustable table. The connection between the two boxes is sealed light-tight with a bicycle tube.\\
	
	\section{Characterization}
	\label{Characterization}
\begin{figure}[h!]
		\includegraphics[width=7.5cm]{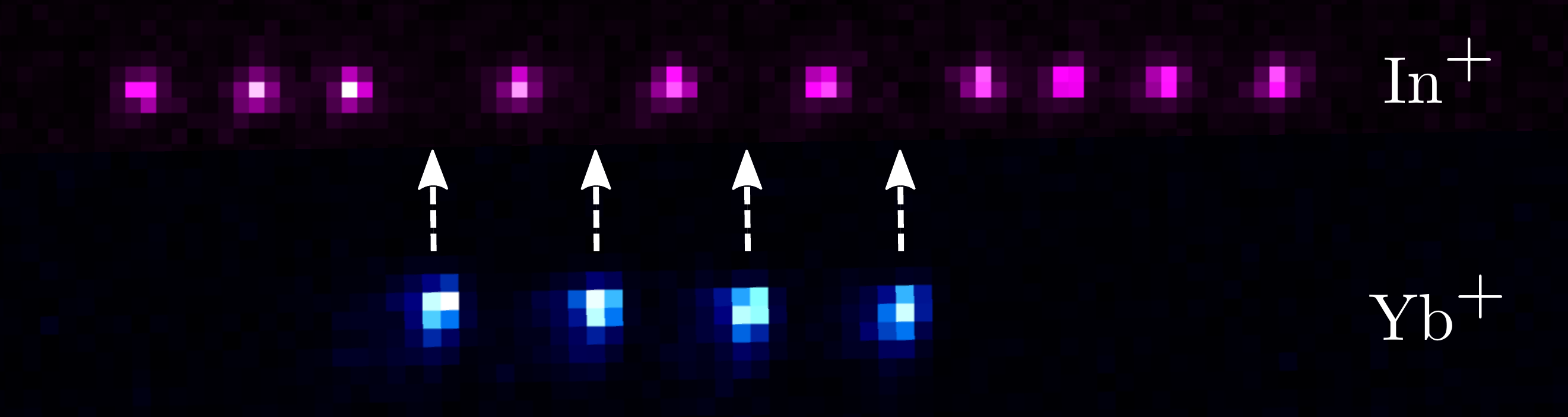}
		\caption{EMCCD image (false color) of a linear Coulomb crystal consisting of 10 In$^+$ and 4 Yb$^+$ ions. The Yb$^+$ ion image is increased to the same magnification as the In$^+$ image and rotated by 1.2\,$^{\circ}$. The simultaneous exposure time is 200\,ms.}
		\label{fig:mixed_crystal}
	\end{figure}

	A false color EMCCD image of a linear 10~In$^+$/4~Yb$^+$ ion crystal is shown in Fig.~\ref{fig:mixed_crystal}. The size of the Yb$^+$ image is increased to match the magnification of the In$^+$ image and is rotated by 1.2\,$^{\circ}$ to compensate a tilt of the system. Due to their different magnifications, the two species are imaged at slightly different positions in vertical direction to avoid overlapping In$^+$ and Yb$^+$ ion positions on the EMCCD camera.\\ 
	
\begin{figure}[h!]
		\includegraphics[width=7.0cm]{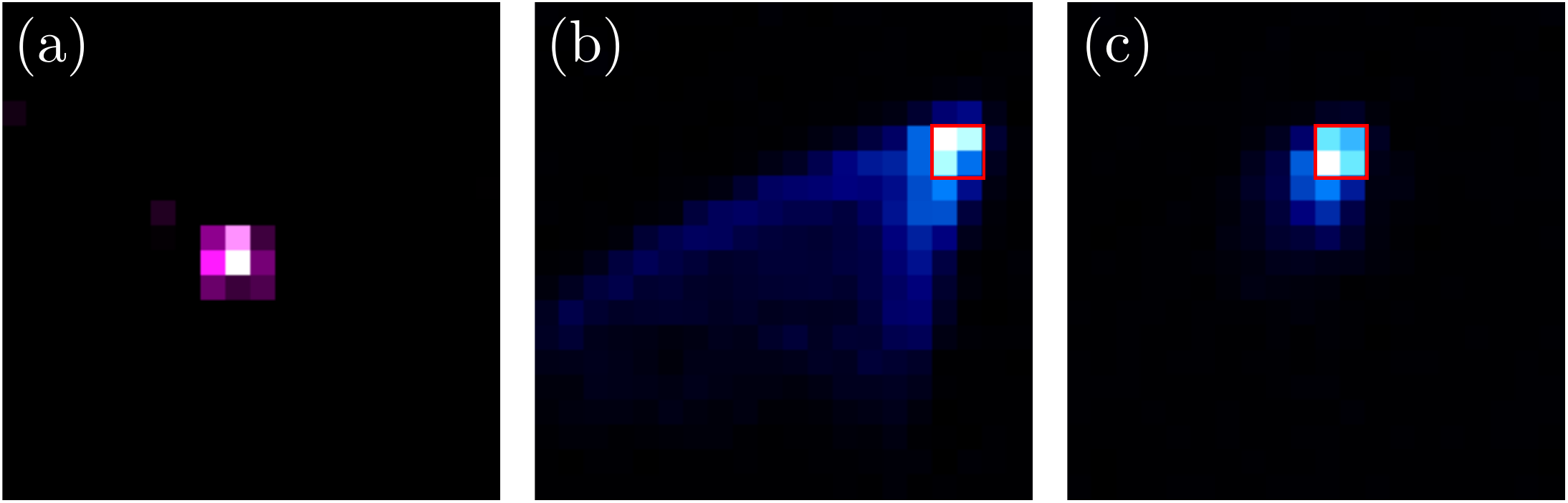}
		\caption{Single ion images: In$^+$ ion image (a) Yb$^+$ ion image with open pinhole (b) and closed pinhole (c). The pinhole cuts away edge rays that lead to the tail in (b). The fluorescence in the indicated red ROI is reduced by 5\,\%. The overall fluorescence summed across the shown image is reduced to one third of (b) in (c).}
		\label{fig:effect_pinhole}
	\end{figure}

	During the assembly of the system, the most critical alignment was the angle with respect to the vacuum window. After alignment optimization, the Yb$^+$ ion image still shows coma on the EMCCD camera as depicted in Fig.~\ref{fig:effect_pinhole} (b). With the closure of the pinhole between the wedged substrate (WS) and the beam combiner (BS) to a diameter of $\approx$\,6\,mm, the fluorescence in the center of the image (indicated by the red square in Fig.~\ref{fig:effect_pinhole} (b) and (c)) is only reduced by 5\,\%. The resulting image is shown in Fig.~\ref{fig:effect_pinhole} (c). The pinhole reduces the $NA$ of the Yb$^+$ beam path onto the EMCCD camera to $\approx$\,0.25, which is accompanied by a reduction of the total fluorescence to 1/3. The In$^+$ ion image shows no indication of coma or other aberrations (see Fig.~\ref{fig:effect_pinhole} (a)).\\
	
\begin{figure}[h!]
		\includegraphics[width=8.5cm]{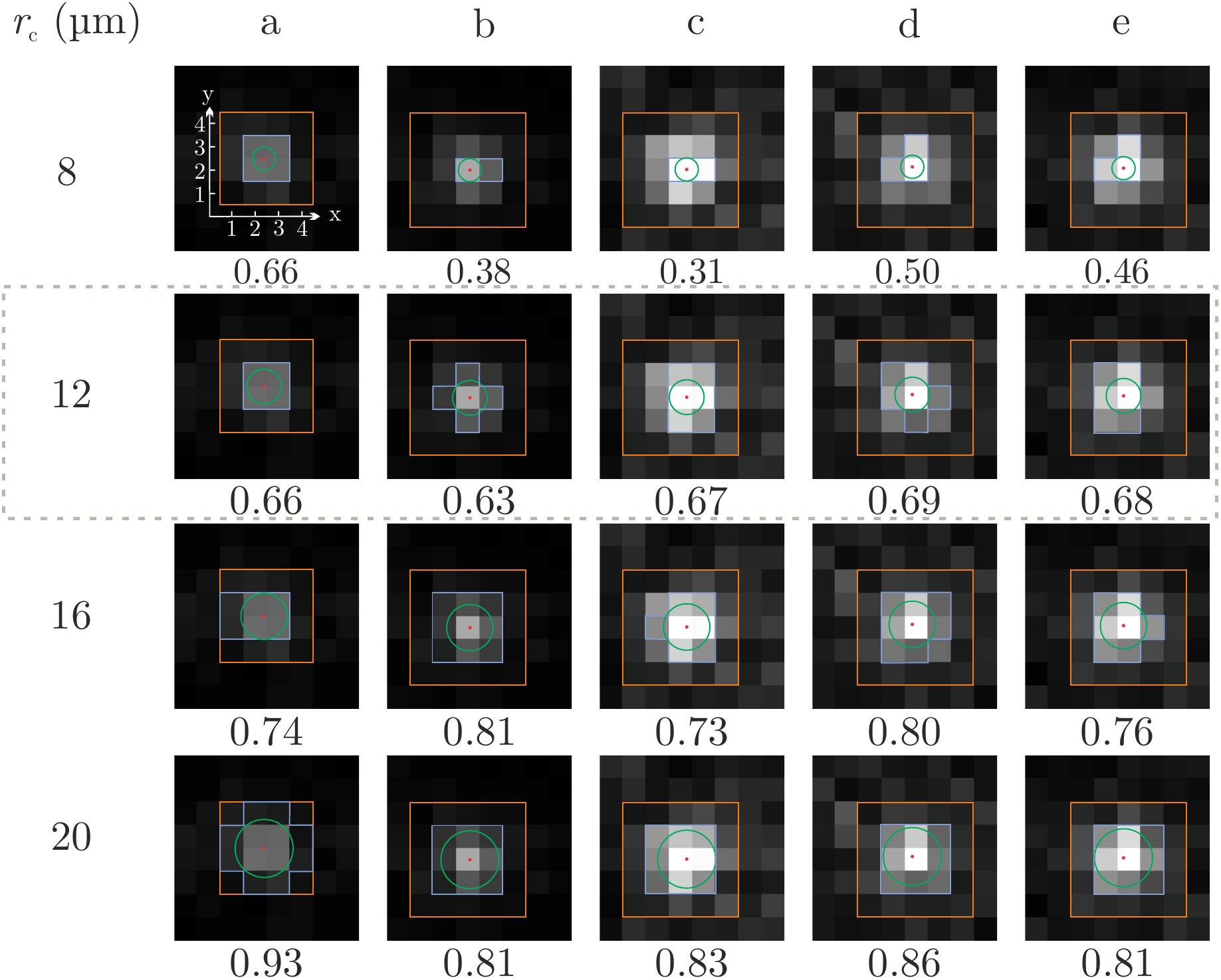}
		\caption{Determination of the radius $r_\mathrm{c,exp}$ which fulfills $f_\mathrm{ee,exp}$\,$\approx$\,0.7 for the 230.6\,nm path. Shown are five images (a-e) of a single In$^+$ ion, recorded with the camera. For the images, the center point (red point) of the ion is calculated via the gray values of the pixels in the orange ROI. The green circle around the center point indicates $r_\mathrm{c}$, which is varied between 8\,$\mu$m and 20\,$\mu$m, and the blue ROI presents the area which has been used to calculate $f_\mathrm{ee,exp}$($r_\mathrm{c}$) given below each image.}
		\label{fig:balance_point_enclosed_fraction}
	\end{figure}

	Fig.~\ref{fig:balance_point_enclosed_fraction} serves to characterize the system performance at 230.6\,nm quantitatively and to compare it to the simulations which consider tolerances described in Sec.~\ref{tolerances}. A possible approach is to determine $f_\mathrm{ee,exp}$($r_\mathrm{c,1p}$\,=\,8\,$\mu$m) from experimental images of individual In$^+$ ions. The large pixel size (16\,$\mu$m\,$\times$\,16\,$\mu$m) of the camera limits the resolution of the evaluation. If the ion is perfectly located in the center of one pixel, $r_\mathrm{c,1p}$ encloses this pixel and the fluorescence in this pixel can be used to calculate $f_\mathrm{ee,exp}$($r_\mathrm{c,1p}$). In most cases, the ion is located such that up to four pixels fall into the range of $r_\mathrm{c,1p}$. For the evaluation in Fig.~\ref{fig:balance_point_enclosed_fraction}, the position of the ion is determined by calculating its center point of brightness within a region of interest (ROI). In the first line of Fig.~\ref{fig:balance_point_enclosed_fraction} (for $r_\mathrm{c}$\,=\,8\,$\mu$m\,=\,$r_\mathrm{c,1p}$), five images of single In$^+$ ions (a-e), recorded with different light intensity and exposure time, are shown. The orange ROI indicates the area in which the fluorescence of the ion is collected and thus considered for the following analysis. A coordinate system ($x,y$) labels the pixels as indicated in the first image (a), where the start ($x$\,=\,0, $y$\,=\,0) is placed in the center of one pixel. The $x$- and $y$-coordinate of the center point are calculated by weighing the gray values with the respective $x$- and $y$-coordinates and the sum of all gray values in the orange ROI. The center points for all images are displayed as red points. Around the center point, a green circle with $r_\mathrm{c}$\,=\,8\,$\mu$m is drawn in the first line of Fig.~\ref{fig:balance_point_enclosed_fraction}. The blue ROI shows which pixels intersect with the green circle and are considered to calculate the value of $f_\mathrm{ee,exp}$($r_\mathrm{c}$) given below each image.\\
	
	The position of the center point varies for the images a-e and 2-4 pixels are taken into account to calculate $f_\mathrm{ee,exp}$($r_\mathrm{c}$), which results in values between 0.31 and 0.66 for the five images. This shows that $r_\mathrm{c}$\,=\,8\,$\mu$m\,=\,$r_\mathrm{c,1p}$ is too small to fulfill $f_\mathrm{ee,exp}$\,$\approx$\,$f_\mathrm{ee,tol}$\,$\approx$\,0.7, derived from the tolerancing of the system. To estimate the value of $r_\mathrm{c}$ which fulfills $f_\mathrm{ee,tol}$\,$\approx$\,0.7, $r_\mathrm{c}$ is increased to 12\,$\mu$m, 16\,$\mu$m and 20\,$\mu$m in the following rows of Fig.~\ref{fig:balance_point_enclosed_fraction} and the evaluation of the images a-e is repeated. The values  $f_\mathrm{ee,exp}$($r_\mathrm{c}$\,=\,12\,$\mu$m) agree with $f_\mathrm{ee,tol}$\,$\approx$\,0.7. Since the blue ROI is larger than the green circle, a too large area is included in the calculation of $f_\mathrm{ee,exp}$, but the pixel size does not allow for better resolution. The evaluations at $r_\mathrm{c}$\,=\,16\,$\mu$m and $r_\mathrm{c}$\,=\,20\,$\mu$m serve to check if $r_\mathrm{c}$ can be increased to $r_\mathrm{c}$\,$>$\,12\,$\mu$m without achieving $f_\mathrm{ee,exp}$\,$>$\,0.7 for some images. For $r_\mathrm{c}$\,=\,16\,$\mu$m, all five values of $f_\mathrm{ee,exp}$\,$>$\,0.7, so for the 230.6\,nm path $r_\mathrm{c,exp}$($f_\mathrm{ee,exp}$\,=\,0.7)\,=\,(12\,$\pm$\,2)\,$\mu$m is concluded. This means that the ratio between the experimentally determined radius $r_\mathrm{c,exp}$ and the radius determined from the toleranced simulation $r_\mathrm{c,1p}$ for $f_\mathrm{ee}$\,=\,0.7 is $r_\mathrm{c,exp}$/$r_\mathrm{c,1p}$\,=\,1.5. \\
	
	For the 369.5\,nm path with $f_\mathrm{ee,tol}$\,$\approx$\,0.6, a detailed characterization with multiple images has been omitted because the image exhibits coma as described earlier. With the pinhole closed, typically 65\% of the fluorescence are collected on 3\,$\times$\,3 pixels which corresponds to $r_\mathrm{c,exp}$\,$\approx$\,24\,$\mu$m.\\
	
	Table~\ref{tab:specs_detection} summarizes relevant parameters of the imaging system for both beam paths. The theoretical detection efficiencies of the PMTs consider the collected fraction of the solid angle, the scattering rate, the scattering pattern, the transmission of the vacuum window and the lenses and the quantum efficiency specified by the vendor.\\
	
	\onecolumngrid
	\begin{center}
		\begin{table}[h!]
			\caption{Specifications of the detection system. The theoretical detection efficiencies for the PMTs are based on the collected fraction of the solid angle, the quantum efficiency of the PMT, the scattering rate and the transmission of the windows and lenses.}
			\label{tab:specs_detection}
			\begin{center}
				\begin{ruledtabular}
					\begin{tabular}{lll} 
						& In$^+$ (230.6\,nm)	& Yb$^+$ (369.5\,nm)\\
						\toprule
						numerical aperture $NA$ & 0.45 & 0.40 (0.25 for the camera)\\
						measured magnification $M$ & 11.64 $\pm$ 0.03 & 9.98 $\pm$ 0.05 (with FBS: 9.88 $\pm$ 0.09)\\
						theoretically collected fraction of the solid angle $\frac{\Omega}{4\pi}$ &0.067 &0.042 (0.016 for the camera)\\
						ratio of measured/theoretical detection efficiency for PMT &	0.6 &0.95\\
						$f_\mathrm{ee,tol}$($r_\mathrm{c,1p}$\,=\,8\,$\mu$m) & 0.7& 0.6\\
						$r_\mathrm{c,exp}$($f_\mathrm{ee,tol}$) ($\mu$m)& 12\,$\pm$\,2 &$\approx$\,24\\
						$r_\mathrm{c,exp}$($f_\mathrm{ee,tol}$)/$r_\mathrm{c,1p}$($f_\mathrm{ee,tol}$) & 1.5 &$\approx$\,3
					\end{tabular}
				\end{ruledtabular}
			\end{center}
		\end{table}
	\end{center}
	\twocolumngrid	
	While for the Yb$^+$ PMT, the achieved fluorescence yield is 0.95 of the expectation, it is only 0.6 for the In$^+$ PMT. A possible reason for the deviation could be a lower quantum efficiency as given in the datasheet of the PMT due to aging processes or larger losses in the vacuum window and mirrors than expected. For the 230.6\,nm path $f_\mathrm{ee,exp}$\,=\,$f_\mathrm{ee,tol}$\,=\,0.7 is reached at a radius $r_\mathrm{c,exp}$\,=\,(12\,$\pm$\,2)\,$\mu$m, which is a factor 1.5 larger than the corresponding radius  $r_\mathrm{c,1p}$\,=\,8\,$\mu$m from the simulation considering the tolerances for the lenses. Important to note is that the In$^+$ image does not exhibit the coma shown in the worst-case scenario in Fig.~\ref{fig:PSF_with_tolerances}~(a). Nevertheless $r_\mathrm{c,exp}$($f_\mathrm{ee,tol}$\,=\,0.7) is larger than expected from this scenario. Most likely the deviation arises from the vacuum window, which has not been included in the tolerancing. A small tilt ($\approx$\,0.2\,\degree) between the whole system and the window as well as a small wedge of the window ($\approx$\,0.02\,\degree) would lead to an image being at least twice as broad as for the ideal system. For the 369.5\,nm path only a rough estimate has been done revealing $r_\mathrm{c,exp}$($f_\mathrm{ee,tol}$\,=\,0.6)\,$\approx$\,24\,$\mu$m which is $\approx$\,3  times larger than the value $r_\mathrm{c,1p}$\,=\,8\,$\mu$m derived from the simulation considering the tolerances for the four lenses. This can only be explained by imperfections of the BS, WS or BC, most likely is a deviation from perfect parallelism or the specified wedge (in case of the WS) for at least one of the components. The alignment angle of the components is not so critical here and can deviate by up to 3\,\degree for the WS, for example, before a deviation would be noticed with the low camera resolution. The observed distortion could be caused by a $\approx$\,\,0.05\,\degree wedge of the BS, which is in the range of the vendors specification.\\
	
	As expected from the tolerances, the imaging of ions on a single pixel is not possible. Nevertheless, the achieved imaging quality is well suited for our purpose of simultaneous spatially-resolved state detection of Yb$^+$ and In$^+$ ions.\\
	
\begin{figure}[h!]
		\includegraphics[width=9.0cm]{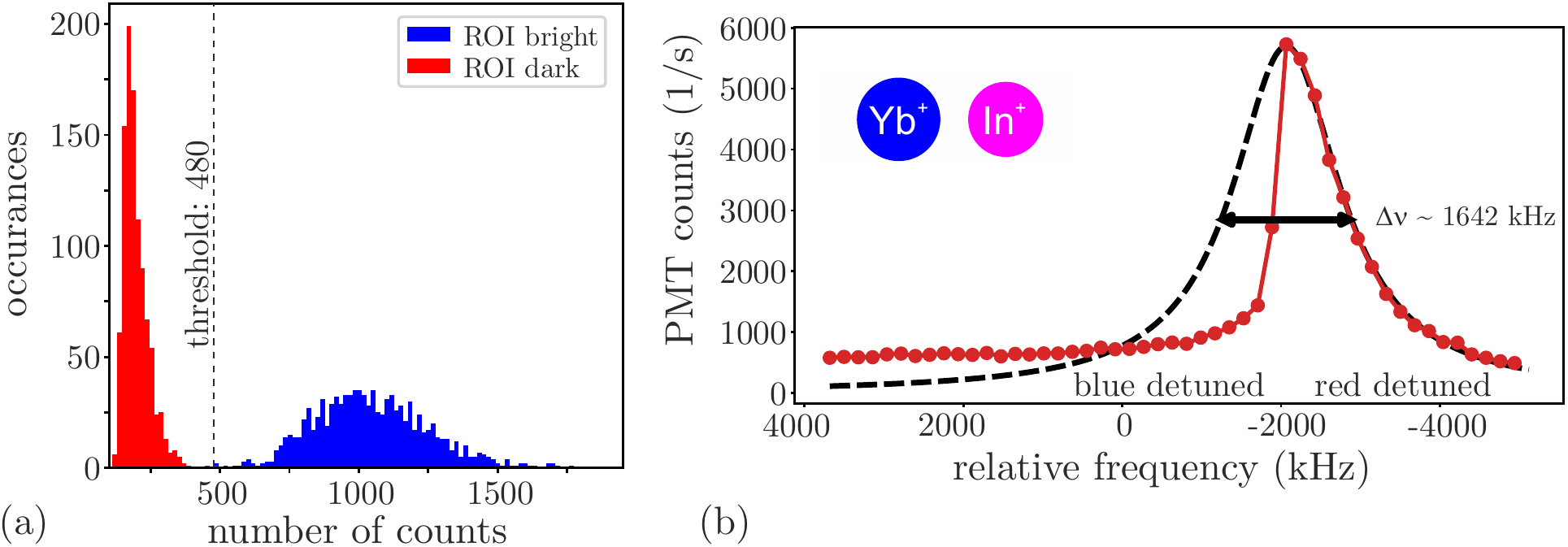}
		\caption{(a) Histogram for In$^+$ state detection with the EMCCD. The detuning from resonance is -$\Gamma$/2, the detection time is 10\,ms, the saturation parameter s\,=\,6, 3\,$\times$\,3 pixels have been binned to one superpixel and is read out as the region of interest (ROI). In red, the dark background counts are shown and in blue the bright counts. (b) PMT recording of a frequency scan across the +9/2~$\rightarrow$~+11/2 Zeeman component of the $^1S_0$~$\leftrightarrow$~$^3P_1$ transition of a single In$^+$ ion at a saturation parameter s\,=\,10. The black dashed line is a Lorentzian fit to the experimental data, indicating a linewidth $\Delta\,\nu$\,=\,1642\,kHz.}
		\label{fig:In_histogram}
	\end{figure}

	Figure~\ref{fig:In_histogram} (a) shows clearly separated histograms for In$^+$ ion state detection with the EMCCD using a detection time of 10\,ms. The fluorescence signal of a single In$^+$ ion that is sympathetically cooled with one Yb$^+$ ion, recorded with the PMT while scanning the frequency across the resonance is illustrated in Fig.~\ref{fig:In_histogram} (b).\\
	
	\section{Conclusion}
	\label{conclusion}
	In summary, we have presented the design of a bichromatic detection system for simultaneous spatially-resolved state detection of dual-species Coulomb crystals in the deep UV on a shared EMCCD camera. To our knowledge this is the first simultaneous detection of dual-species Coulomb crystals on one EMCCD camera. The system is placed outside of the vacuum chamber and features an $NA$\,=\,0.45 for imaging of In$^+$ (230.6\,nm) ions on both the PMT and camera. For Yb$^+$ (369.5\,nm) ions the system provides an $NA$\,=\,0.40 on the PMT and $NA$\,=\,0.25 on the camera. The introduced concept can be adapted to other dual-species ion combinations. If an $NA$\,$>$\,0.4 is desired for the transmitted wavelength (Yb$^+$) the beamsplitter, beam combiner and the wedged substrate need to be optimized and considered in the tolerancing. \\
	
	With the main goal of building an In$^+$ multi-ion frequency standard \cite{Keller_2019_PRA}, the simple pinhole approach was chosen for the Yb$^+$ beam path to the camera. For the state detection of In$^+$ ions we improved the detection time by a factor 4-6 compared to previous work, where no spatial resolution and bichromatic detection was realized \cite{Becker_2001, Ohtsubo_2019}.
	
	\begin{acknowledgments}
		We acknowledge funding by the Deutsche Forschungsgemeinschaft (DFG) through grant CRC 1227 (DQ-mat, project B03) and through Germany’s Excellence Strategy EXC2123 QuantumFrontiers. We thank department 5.5 for the machining of the mechanical mounting of the setup and especially M. M{\"u}ller for its design. Furthermore, we thank J.~Kiethe and J.~Keller for helpful discussions and comments on the manuscript.
	\end{acknowledgments}

\section*{Data availability}
The data that support the findings of this study are available from the corresponding author upon reasonable request.
%

\end{document}